\documentclass[twocolumn,secnumarabic,amssymb]{revtex4}
\usepackage{graphicx}
\usepackage{amsmath}
\usepackage{amssymb}
\usepackage{dcolumn}
\usepackage{bm}

\begin{document}

\title{The interpretation of broad diffuse maxima using superspace crystallography}

\author{Ella Mara Schmidt}
\email{ella.schmidt@fau.de}
 
\author{Reinhard B. Neder}%
\affiliation{Institute for Crystallography and Structural Physics, Friedrich-Alexander Universit\"at Erlangen N\"urnberg.}

\date{\today}

\begin{abstract}
Single crystal diffuse scattering is generally interpreted using correlation parameters that describe probabilities for certain configurations on a local scale. 
In this paper we present a novel interpretation of diffuse maxima using a disordered superspace approach. 
In $(D+d)$-dimensional superspace two modulation functions are disordered along the superspace axis $\bm{a}_{s,i}$ for $i = 1,..,D$, while the periodicity along the internal dimensions is maintained.
This simple approach allows the generation of diffuse maxima of any width at any position in reciprocal space.
The extinction rules that are introduced by superspace symmetry are also fulfilled by the diffuse maxima from structures generated using the disordered superspace approach.
In this manuscript we demonstrate the disordered superspace approach using a simple a simple two-dimensional binary disordered system.
The extension of the approach to $(3+D)$-dimensional superspace is trivial. 
The treatment of displacement and magnetic disorder in a similar approach is straight forward.

\end{abstract}

\maketitle

\section{Introduction}
Diffuse scattering has been observed in single crystal diffraction experiments since the beginning of X-ray crystallography \cite{Laue1918}.
For moderately complex structures the average structure determination using Bragg scattering has long been a routine method and long range ordering patterns can be easily identified.
More recently the solution and refinement of modulated crystal structures has advanced \cite{petvrivcek2014crystallographic,janssen2014aperiodic,janssen2018aperiodic,van2007incommensurate}.
The superspace approach made the interpretation of modulated structures straight forward, again opening up the possibility to identify long range albeit aperiodically ordered patterns.
The interpretation of diffuse scattering on the other hand generally relies on a set of short range order parameters \cite{keen2015crystallography, welberry2016one}. 
Compared to the analysis of long-range ordered structures, many parameters are needed to describe a diffuse scattering pattern.

In contrast to long range ordered aperiodically modulated structures, there are almost no widely applicable programs that perform structure solution for diffuse scattering.
Instead a system specific approach is taken for most systems that are investigated in terms of correlated disorder.
So far no standard algorithm has been established to solve any arbitrary disordered structure.

In the early beginnings of diffuse scattering analysis, computational power was barely available. 
Therefore, it was impossible to build large computer based models for the calculation of diffuse scattering patterns. 
This required the deviation of analytical models to describe diffuse scattering \cite{warren1951atomic, jagodzinski1949eindimensionale, hayakawa1975equations}.
The analytical description has lately been used to successfully identify short range order patterns \cite{burgi2005stacking,schmidt2017diffuse} and is in terms of computational modelling by far the most efficient approach to take.

The 3d-$\Delta$PDF method employs three dimensional pair distribution functions for the diffuse scattering analysis and can be seen as an extension of the analytical models \cite{weber2012three}.
This method comes along with a powerful software (Yell \cite{simonov2014yell}) and the interpretation of short range order phenomena in moderately complex systems is straight forward.

The most widely used approach for refining diffuse scattering patterns is based on Monte-Carlo methods. 
There are two different approaches for Monte-Carlo modelling: direct and reverse Monte-Carlo modelling \cite{neder2008diffuse}.
The direct modelling approach requires an initial disorder model.
Based on such an initial idea, powerful programs, such as DISCUS \cite{neder2008diffuse}, generate a disordered crystal structure that obeys the short range order parameters from the model.
The calculated diffuse scattering is then compared with the experimental results 
and the short range order parameters are adapted until sufficient agreement is reached. Recent applications include e.g. \cite{welberry2014diffuse, burgi2005supramolecular}.
The advantage of direct Monte-Carlo modelling is the direct interpretation of physically meaningful short range order parameters; but comes at the cost of a high computation time.
The large setback is the initial model needed for the short range order, which is in general not straight forward but relies on expert knowledge.  
Many iterations are possibly needed until a meaningful set of short range order parameters is found, that describes the diffuse scattering at hand.

Reverse Monte-Carlo modelling is also widely applied in diffuse scattering analysis. 
In general a big crystal is built and then the atoms or molecules within this crystal are displaced or exchanged until a calculated diffuse scattering pattern agrees with the measured intensities.
The big advantage is that no initial model, and expert guess is needed but the resulting crystal structure is not straight forward to interpret in terms of correlation coefficients \cite{proffen1997analysis}. 
Reverse Monte-Carlo modelling has been successfully applied to powder diffraction data since its introduction in 1988 \cite{mcgreevy1988reverse}.

All the approaches described above have in common that the interpretation of the diffuse scattering is based on a set of correlation coefficients. 
Ray Withers mainly shaped the interpretation of diffuse scattering in terms of modulation waves \cite{withers2015modulation, welberry1995modulation}.
The diffuse scattering is interpreted as the result of many modulation waves acting on the basic crystal structure. 
The modulation wave is assumed to have an arbitrary origin within the structure and to generate broad diffuse maxima, the amplitude is damped, e.g. in a Gaussian fashion, as the wave diverges from this origin.
While this interpretation significantly differs from the correlation coefficient approach it still needs many parameters to describe a disordered structure (e.g.  \cite{welberry1995modulation} uses $3\times10^{5}$ perturbations to create a structure that agrees with the experimentally observed diffuse scattering).
 
This paper presents a different modulation wave approach, using the tools of superspace crystallography.  
The introduction of disorder into superspace requires only few parameters to place a diffuse maximum at a freely chosen position in reciprocal space and with freely chosen width.
The approach presented here leaves the internal dimension perfectly periodic and introduces phase domains $A$ and $B$ in the remaining superspace dimensions. The width of the diffuse maximum is inversely proportional to the size of the phase domains in superspace. This relationship is completely analogous to the inverse relationship between diffuse maxima and domain sizes in the case of short range ordered materials.

\section{Superspace and modulation functions}
Bragg reflections in reciprocal superspace are indexed with $(3+d)$ integers. 
The main Bragg reflections form the reciprocal lattice $\bm{G}$ using the three reciprocal lattice vectors $\bm{a}_{1}^{\ast}$, $\bm{a}_{2}^{\ast}$ and $\bm{a}_{3}^{\ast}$. 
All main Bragg reflections can be indexed using the lattice vector $\bm{H} = (h \: k \: l)$, with integer $h$, $k$ and $l$. 
Satellite reflections at $\bm{H} + \sum_{j=1}^{d} m_{j}\bm{q}_{j}$ with integer $m_{j}$ appear in addition to the main reflections, with the modulation wave vectors $\bm{q}_{j}$:
\begin{equation}
\bm{q}_{j} = \sum_{i=1}^{3} \sigma_{i,j}\bm{a}_{i}^{\ast}
\end{equation}
For crystals with one modulation vector, the fourth reciprocal lattice direction $\bm{a_{4}^{\ast}}$ is chosen orthogonal to physical 3D reciprocal space with the length of $\bm{a}_{4}^{\ast} = \bm{q}$, which ensures that the satellite reflections can also be indexed with $(3+1)$ integer indices.
For $(3+d)$ dimensional superspace, the three dimensional reciprocal space is then seen as the projection of a $(3+d)$ dimensional reciprocal superspace \cite{van2007incommensurate}.
If more than one modulation wave exists, all $d$ additional reciprocal axes are likewise orthogonal to three dimensional physical space.

The Fourier transform of the reciprocal superspace lattice yields a $(3+d)$ dimensional direct lattice.
While in conventional three dimensional crystallography a point atom approximation is taken, atoms are transformed into wavy strings, that are on average parallel to the additional dimensions in superspace \cite{van2007incommensurate}. 
Direct three dimensional space is obtained as a cut through superspace perpendicular to the additional dimensions. 
The intersection of this cut with the atom surfaces in superspace, yields the value of the modulation function at this lattice position in real space.

\begin{figure*}
	\includegraphics{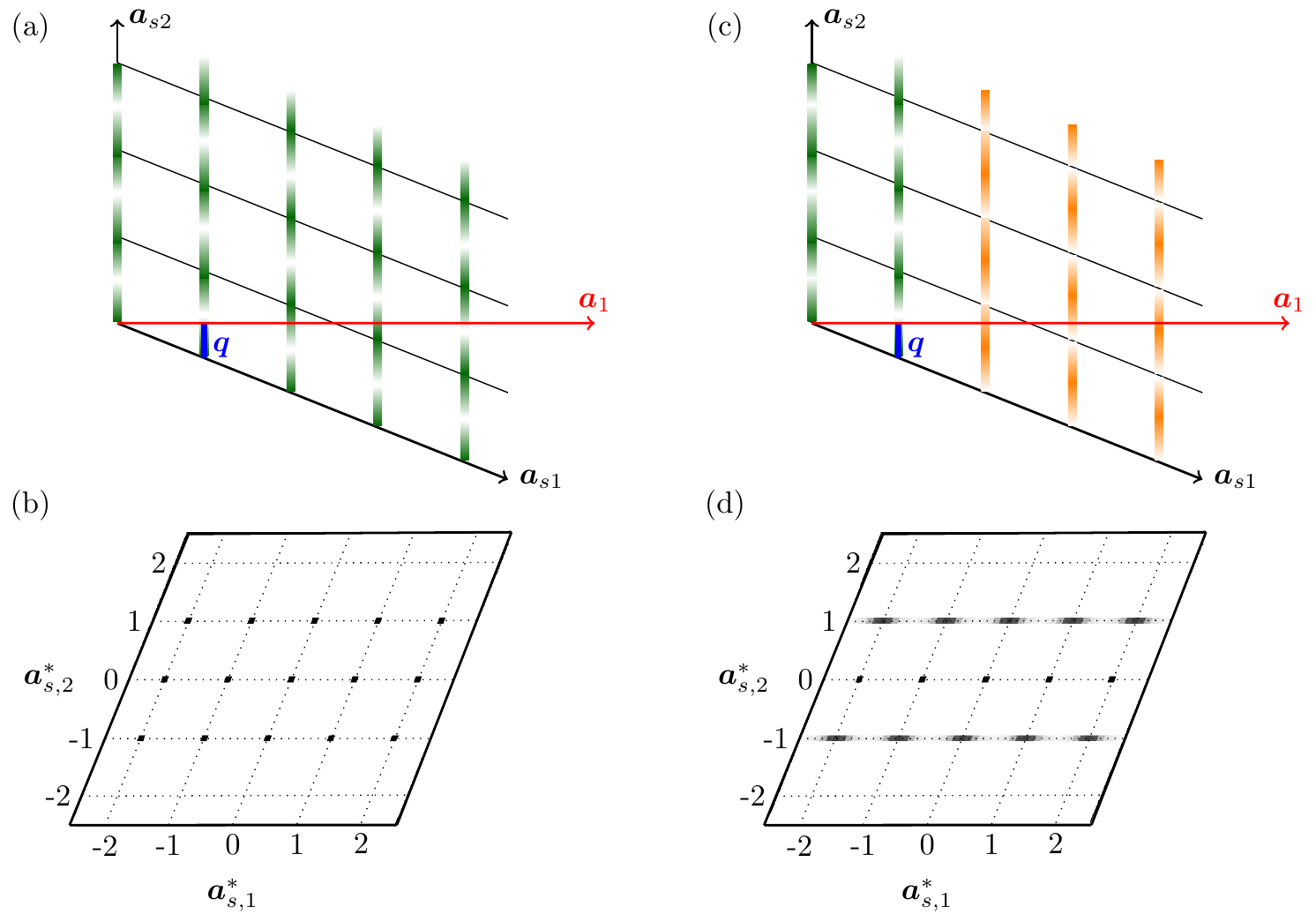}
	\caption{(a)$(1+1)$-dimensional ordered superspace for a one dimensional basic structure with an occupational modulation. The probability to find an atom at site $x$ is indicated by the colour transparency. (b) Reciprocal $(1+1)$-dimensional ordered superspace. For occupational modulations only first order satellites are observed, i.e. in $(1+1)$-dimensional reciprocal superspace only  Bragg reflections at $0\bm{a}_{s,2}^{\ast}$ and $\pm1\bm{a}_{s,2}^{\ast}$ are observed.
	(c)$(1+1)$-dimensional disordered superspace for a one dimensional basic structure with an occupational modulation. Two types of modulation functions $p_{+}(x)$ (green) and $p_{-}(x)$(orange) are introduced. The modulation functions are phase shifted by $\pi$ with respect to each other. (d) Reciprocal $(1+1)$-dimensional disordered superspace. The sharp satellite reflections at $\pm1\bm{a}_{s,2}^{\ast}$ are transformed into broad diffuse maxima.}
		\label{fig:SuperSpace}
\end{figure*}

For a a simple occupational modulation with modulation wave vector $\bm{q}$ the occupancy of an atom $j$ in unit cell at vector $\bm{n}$ may be given as \cite{janssen2014aperiodic}:
\begin{equation}
p(\bm{n} + \bm{r}_{j}) = \bar{p} +  A\cos(2\pi\bm{q}(\bm{n} + \bm{r}_{j}) + t)
\label{eq:Mod1}
\end{equation}
$\bar{p}$ is the average occupancy.
The amplitude $A$ of the modulation function has to be chosen so that $0 \le p(\bm{n} + \bm{r}_{j}) \le 1$ is ensured. 
The term $t$ determines the initial phase of the modulation function.
The ordered superspace model is depicted in Figure \ref{fig:SuperSpace}(a) for a one dimensional crystal in two dimensional superspace.

\section{Disorder diffuse scattering}
The origin of diffuse scattering can be related to static or dynamic disorder. 
While thermal diffuse scattering accounts for effects that are connected to the correlated thermal motion of atoms, disorder diffuse scattering covers effects due to static occupational or displacive alterations of the crystal structure, that are only present on a local scale.
Here we consider disorder diffuse scattering.

Our initial system that is used to illustrate the concepts of the super space approach to diffuse scattering is a binary disordered system, that only exhibits occupational disorder.
Two atom types $A$ and $B$ are distributed over the lattice points.
The diffuse intensity $I_{D}(\bm{h})$ of such a system is given as \cite{schmidt2017diffuse}:
\begin{equation}
\begin{split}
I_{D}(\bm{h}) = &Nm_{A}m_{b}\left|f_{A}(\bm{h}) - f_{B}(\bm{h}) \right|^{2} \\
& \left[1 + \sum_{\vec{v} \in V_{p}} \alpha_{\bm{v}} \cos(2\pi\bm{h}\bm{v})\right] 
\end{split}
\label{eq:SRO}
\end{equation}
$N$ is the number of atoms in the crystal, $m_{A}$ and $m_{B}$ are the relative abundances of species $A$ and $B$. 
$f_{A}$ and $f_{B}$ are the atomic form factors of species $A$ and $B$. 
The sum runs over all interatomic vectors $\bm{v}$ in the positive real vector half-space:
\begin{equation}
\begin{split}
V_{p} = \Bigl\{ &\bm{v} = \left. \left( u,v,w\right)^{T} \right\vert \left(u>0  \right) \lor \left(u=0 \land v > 0 \right) \Bigr.\\
 &\Bigl. \lor \left(u=v=0 \land w > 0 \right) \Bigr\}
\end{split}
\label{eq:Vp}
\end{equation}
$\alpha_{\bm{v}}$ is the Warren-Cowley short range order parameter \cite{warren1951atomic}:
\begin{equation}
\alpha_{\bm{v}} = 1 - \frac{p^{AB}_{\bm{v}}}{m_{A}m_{B}}
\end{equation}
where $p^{AB}_{\bm{v}}$ is the probability to find a species $B$ at a vector $\bm{v}$ from a species $A$.
For random distribution of species $A$ and $B$ $\alpha_{\bm{v}} = 0$ for all interatomic vectors $\bm{v} \ne (0,0,0)^{T}$ and only monotonic diffuse Laue scattering is observed.

\begin{figure}
	\includegraphics[width = 0.4\textwidth]{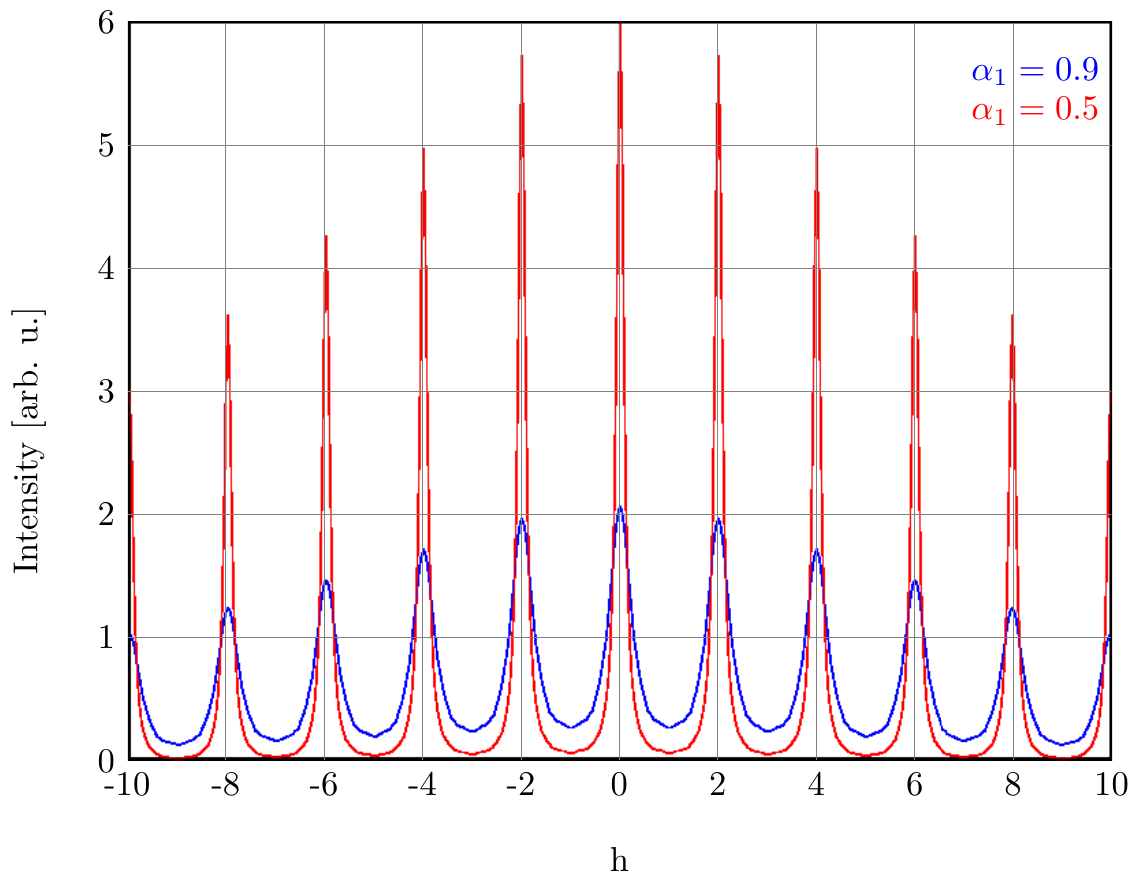}
	\caption{\label{fig:Diffuse1d} Diffuse scattering of an one dimensional Au:Ag alloy with occupational disorder. The calculated diffuse scattering is demonstrated for two different first neighbour Warren-Cowley short range order parameters $\alpha_{1} > 0 $. The diffuse scattering was calculated using the DISCUS program \cite{neder2008diffuse}, for X-ray scattering. The Bragg intensities were omitted in the calculation.}
\end{figure}

In a system that is positively correlated, likewise species cluster together and form domains.
In such a system all $\alpha_{\bm{v}}$ for  $\bm{v}$ that are within the average domain size, are positive and the diffuse scattering accumulates around the Bragg reflections.
Figure \ref{fig:Diffuse1d} shows the diffuse scattering for two different first order correlation parameters $\alpha_{1}$.

\section{Disordered superspace - A one dimensional example}
The following example illustrates the concept of disordered superspace using a binary alloy. To simplify the initial description, we use a (1+1)-dimensional example with a single atom position in the average unit cell.
For a structure with sharp satellite reflections, (3+d)-dimensional superspace is perfectly periodic. In the case of an occupational modulation, the atom strings in superspace are density modulations along straight lines parallel to the axis $\bm{a}_{s,2}$, with a modulation function:
\begin{equation}
p(x) = \bar{p} + A\cos(2\pi qx) 
\end{equation}
The amplitude A of the modulation function must ensure that the probability $p(x)$ is limited to the interval [0,1].
Equivalently to a perfectly periodic crystal in 3D, the Fourier transform of the periodic (3+d) structure yields reciprocal space, where the intensity is zero except for Bragg reflections. Here the Bragg reflections include the Bragg reflections of the average structure and the satellite reflections. If the density modulation in superspace is a simple cosine wave, the intensity of satellite reflections is zero except for first order satellites with $m = \pm 1$ (see Figure \ref{fig:SuperSpace}(b)). 

To describe broad diffuse maxima, the periodicity of superspace must be disrupted. To do so we introduce two different modulation functions: 
\begin{align}
p_{+}(x) &= \bar{p} + A\cos(2\pi qx) \\
p_{-}(x) &= \bar{p} - A\cos(2\pi qx) 
\end{align}
There are no experimental means of distinguishing these two modulation functions.
The two modulation functions are distributed along $\bm{a}_{s,1}$. At each position perfect order is maintained  parallel to $\bm{a}_{s,2}$, see Figure \ref{fig:SuperSpace}(c). The disorder in superspace along the $\bm{a}_{s,1}$ axis manifests itself as diffuse scattering in rods parallel $\bm{a_{s,1}^{\ast}}$ at the position of the satellite reflections at integer positions along $\bm{a_{s,2}^{\ast}}$. The two modulation functions are identical except for a shift by $1/2 \bm{a}_{s,2}$. Thus, in projection onto the axis $\bm{a}_{s,1}$ the structure appears periodic and no diffuse scattering is observed at the position of the main Bragg reflections. As each of the shifted modulation functions in itself is strictly periodic along $\bm{a}_{s,2}$, the diffuse scattering is limited to integer positions along the $\bm{a_{s,2}^{\ast}}$ axis (see Figure \ref{fig:SuperSpace}(d)). 

Physical reciprocal space is a projection of reciprocal superspace onto three dimensional space. In this $(1+1)$-dimensional example, it consists of the sharp Bragg reflections of the average structure plus the projected diffuse intensity. The exact distribution of the diffuse intensity depends on the ordering scheme for the two modulation functions in superspace.

\begin{figure}
	\includegraphics[width=0.4\textwidth]{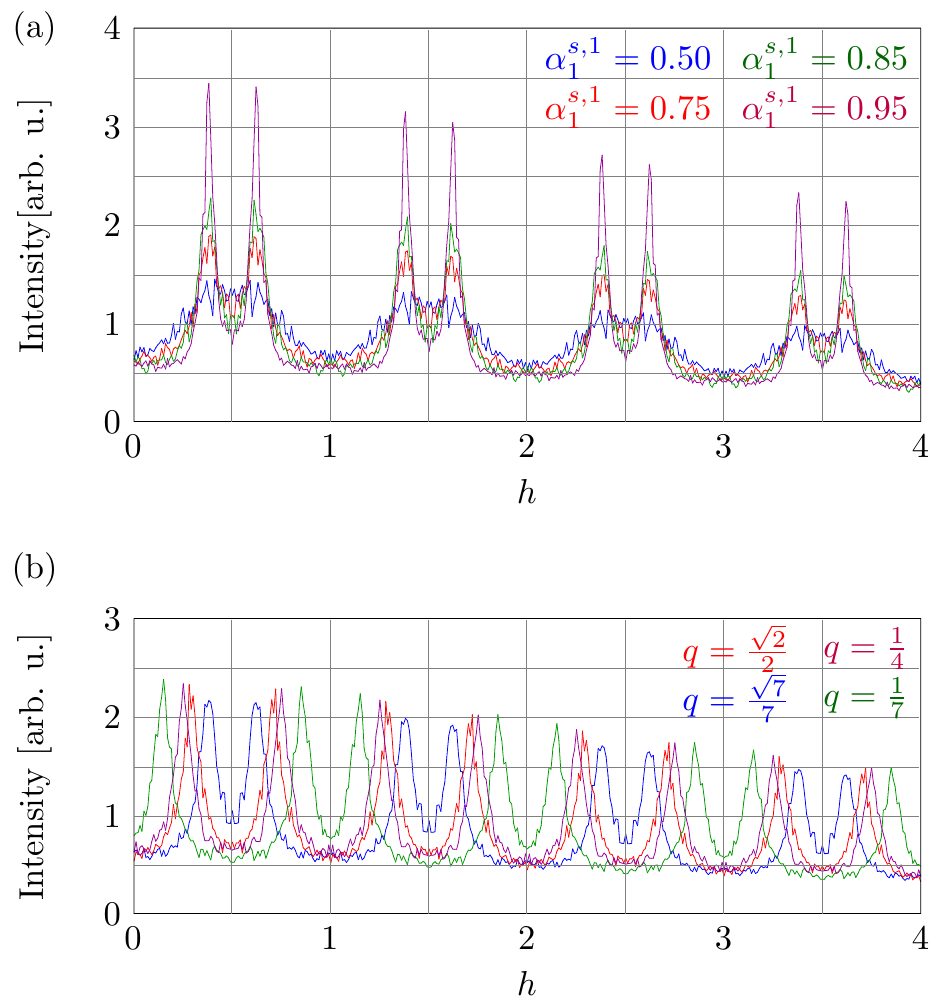}
	\caption{\label{fig:DiffuseSatellite1d} (a) Diffraction pattern for a disordered superspace model showing occupational modulation.
		A one dimensional AuAg 1:1 crystal is simulated. The modulation functions $p_{+}(x) = 0.5 + 0.5\cos(2\pi qx)$ and $p_{-}(x) = 0.5 - 0.5\cos(2 \pi qx)$ are used. Several different Warren-Cowley short range order parameters $\alpha_{1}^{s,1}$ are used. $q=\frac{\sqrt{7}}{7}$.
		(b) Same simulation as in (a); the effect of several different modulation wave vectors is shown. $\alpha_{1}^{s,1} = 0.85$. For computational details see Appendix.}
\end{figure}

A direct space realization can be modeled by analyzing the value of each modulation function at a suitable intersection of superspace with the physical direct space. If the modulation wave vector is of incommensurate length, any cut at a position $t$ along the internal axis, in this $(1+1)$-dimensional example along $\bm{a}_{s,2}$ will give an equivalent structure. If at any position at the cut of $\bm{a}_{1}$ with superspace the modulation function $p_{+}(x)$ is encountered, its value gives the probability to place an atom of type A into the physical structure. If the modulation function $p_{-}(x)$ is encountered, its value gives the probability to place an atom of type A into the physical structure. The diffuse intensities in Figure \ref{fig:DiffuseSatellite1d} are calculated from examples of these one dimensional Au:Ag 1:1 structures. As both atom species are at equal probability, m$_A$=m$_B$ the modulation functions are:
\begin{align}
p_{+}(x) = 0.5 + 0.5\cos(2\pi qx) \\
p_{-}(x) = 0.5 - 0.5\cos(2 \pi qx)
\end{align}  
The diffuse intensity distribution calculated for these truly one-dimensional structures corresponds exactly to the projection of the diffuse intensity in (1+1)-dimensional reciprocal space and is equivalent to the diffuse intensity obtained for the one-dimensional structures used to calculate the diffuse intensity in Figure \ref{fig:Diffuse1d}. 
Variation of the Warren-Cowley short range order parameters $\alpha_{1}^{s,1}$ gives control over the width of the diffuse maxima, while their locations are defined by the length of the modulation wave vector $\bm{q}$. 

In Figure \ref{fig:DiffuseSatellite1d} the Warren-Cowley short range order parameters have been limited to the range [0,1]. This range is sufficient, as it allows to vary the width of the diffuse maxima from sharp satellites for $\alpha=1$ to a completly flat diffuse band for $\alpha=0$. A negative short range order paramter would shift the location of the diffuse maxima in reciprocal superspace by $\frac{1}{2} \bm{a}_{s,1}^{\ast}$, see Figure \ref{fig:SuperSpace}. After the projection of reciprocal superspace onto physical superspace the diffuse maxima would be located at a different q-vector. This shift is, however, preferably performed by changing $\bm{q}$, as demonstrated in \ref{fig:DiffuseSatellite1d}b. 
The computational procedure for model building is described in the Appendix.

\section{(2+1)D Disordered superspace - Diffuse lines and extinction conditions}

\begin{figure}
	\includegraphics{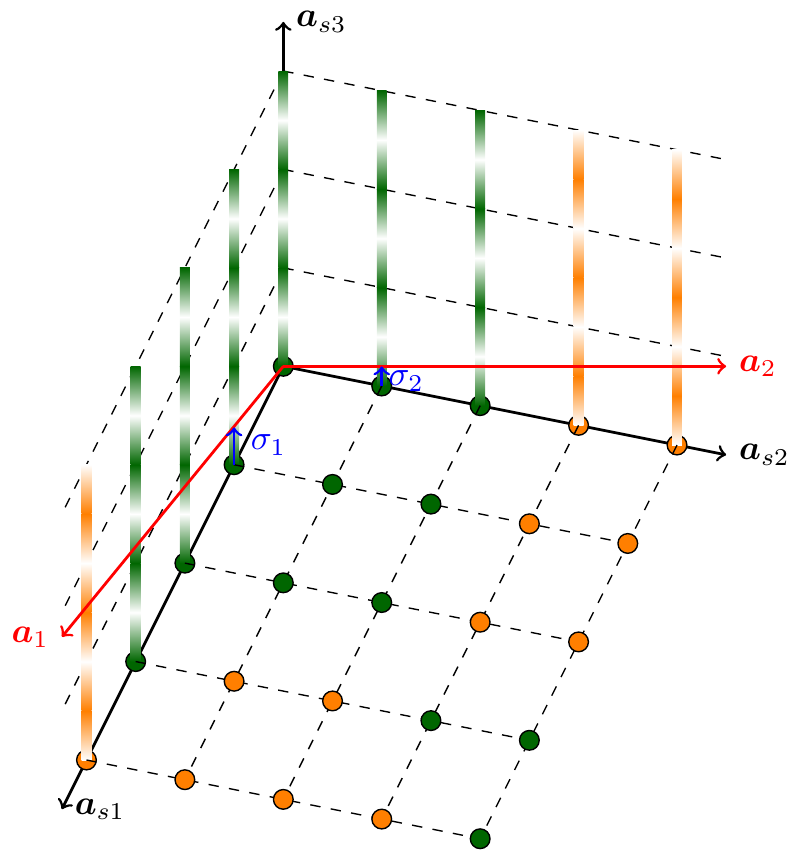}
	\caption{\label{fig:DiffuseSatellite2d} Schematic drawing of a (2+1)-dimensional superspace showing an occupational modulation.
		The two modulation functions $	p_{+}(\bm{x}) = 0.5 + 0.5 \cos(2\pi\bm{q}\bm{x}) $ and $	p_{-}(\bm{x}) = 0.5 - 0.5 \cos(2\pi\bm{q}\bm{x}) $ are indicated in green and orange. The components of the modulation wave vector $\bm{q} = (\sigma_{1}, \sigma_{2})$ define the tilt of the superspace layer defined by $\bm{a}_{s,1}$ and $\bm{a}_{s,2}$ versus physical space, spanned by $\bm{a}_{1}$ and $\bm{a}_{2}$.}
\end{figure}

The concept described in the previous section can easily be expanded to two or three dimensions. Unless restricted by symmetry \cite{van2007incommensurate}, the satellite vector, respectively the diffuse maxima may possess arbitrary coordinates in reciprocal space. 
The ordering of the phase domains is equivalent to the one dimensional case, but the phase domains may have different sizes along different superspace directions. 
Figure \ref{fig:DiffuseSatellite2d} shows a schematic representation of a $(2+1)$-dimensional superspace.

\begin{figure*}
	\includegraphics{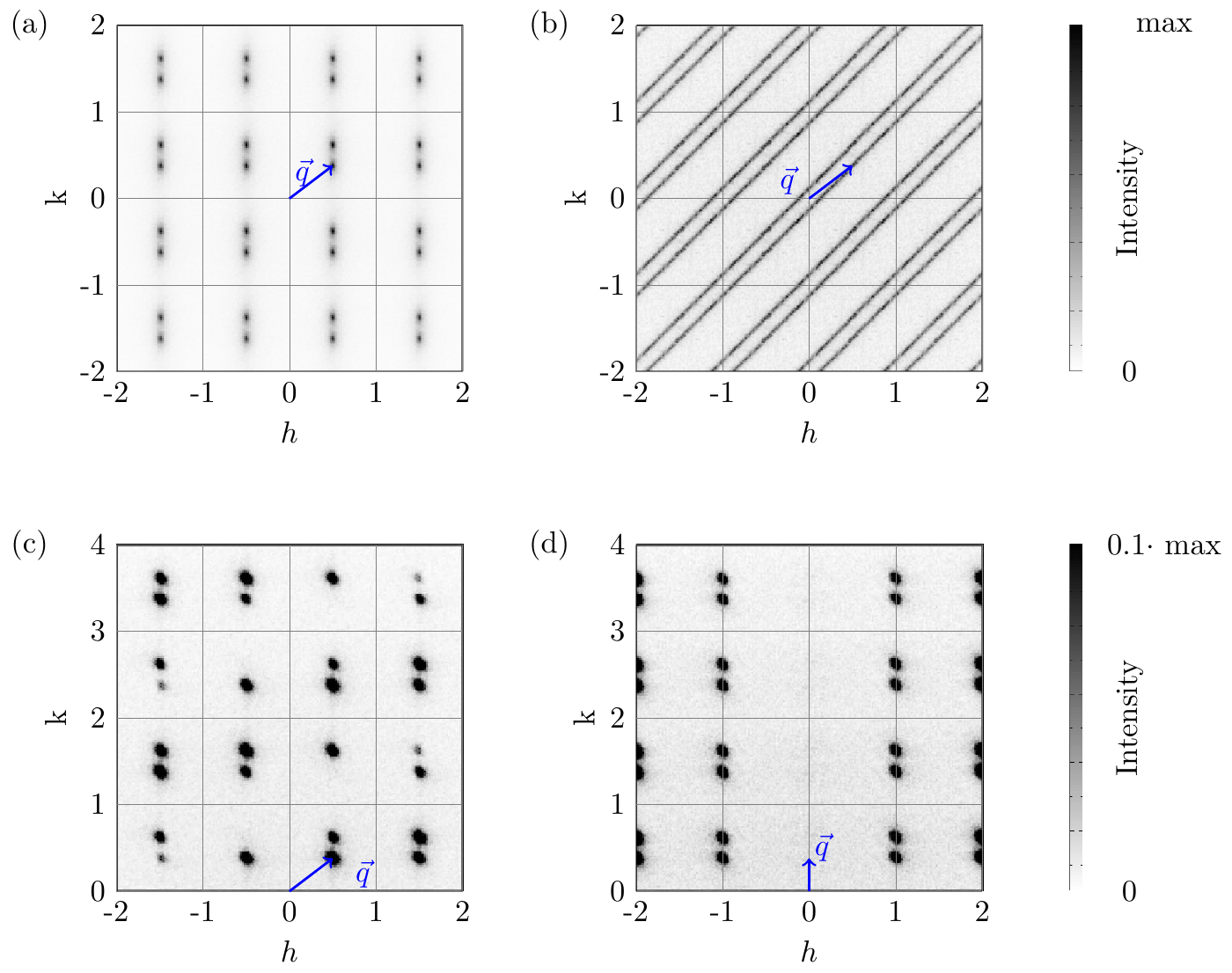}
	\caption{\label{fig:DiffuseSatellite2d_2} Diffraction pattern of (2+1)-dimensional disordered superspace models. Primitive unit cells in $p1$ with only one atom per unit cell are used in (a) and (b), with a modulation wave vector $\bm{q} = \left(\frac{1}{2}, \frac{\sqrt{7}}{7}\right)$. 
		(a)	$\alpha^{s}_{(1,0)} = 0.7$, $\alpha^{s}_{(0,1)} = 0.5$.
		(b) $\alpha^{s}_{(1,1)} = 0$, $\alpha^{s}_{(\bar{1},1)} = 0.95$.
		(c) The diffuse satellite reflections follow the main reflection extinction condition of tier respective parent main reflection $0k: \quad k=2n$. Superspace group $pg1\left(\frac{1}{2}\beta\right)$, $\bm{q} = \left(\frac{1}{2}, \frac{\sqrt{7}}{7}\right)$.
		(d) The diffuse satellite reflections also follow the extinction conditions for internal symmetry. Superspace group $pm1\left(0\beta\right)s0$, $\bm{q} = \left(0, \frac{\sqrt{7}}{7}\right)$.  }
	\end{figure*}

The simulated diffraction patterns shown in Figure \ref{fig:DiffuseSatellite2d_2} are generated from two dimensional AuAg model crystals with occupational disorder.
The width of the diffuse maxima in two dimensional reciprocal space is defined by the Warren-Cowley short range order parameters along $\bm{a}_{s1}$ and $\bm{a}_{s2}$, see Figure \ref{fig:DiffuseSatellite2d_2}(a).
In this example the satellite reflections are broader along $k$ than along $h$ since $\alpha^{s}_{(1,0)} < \alpha^{s}_{(0,1)}$.

The ordering pattern in superspace allows a straight forward generation of diffuse rods in reciprocal space; as is demonstrated in Figure \ref{fig:DiffuseSatellite2d_2}(b).
The modulation functions are almost perfectly ordered along the superspace direction $(\bar{1},1)$ ($\alpha^{s}_{(\bar{1},1)} = 0.95$), while along $(1,1)$ the modulation functions are not ordered ($\alpha^{s}_{(1,1)} = 0$). Note that the choice of the satellite vector $\bm{q}$ is independent of the directions along which superspace is ordered. The specific shape of the diffuse maxima is defined by directions along which superspace is subject to short range order.
The generation of diffuse rods as described here can be directly applied to the diffuse scattering in ThAsSe as described in \cite{withers2004low}, where the diffuse rods run along $\bm{H} \pm \approx 0.14\langle 110 \rangle ^{\ast} \pm \beta \langle 1\bar{1}0 \rangle ^{\ast} \pm \gamma [001]^{\ast}$, with $\beta$ and $\gamma$ essentially continuous.

Extinction conditions in single crystal diffuse scattering contain valuable informations \cite{withers2015modulation}.
In conventional crystallography reflection conditions are encountered whenever the basic structure obeys symmetry operations with a translational component, such as lattice centring, glide planes, or screw axes.
In superspace crystallography, the same reflection conditions are encountered for the main Bragg reflections and their associated satellites \cite{van2007incommensurate}, when the basic structure shows symmetry.
The superspace groups as listed in the \emph{International Tables for Crystallography C} \cite{bryan1993international}, additionally allow a symmetry transformation in superspace, yielding further reflection conditions for the satellite reflections.

To demonstrate the extinction conditions for the diffuse satellite reflections, model structures in super space groups  $pg1\left(\frac{1}{2}\beta\right)$ and $pm1\left(0\beta\right)s0$ are generated (see Figure \ref{fig:DiffuseSatellite2d_2}(c) and (d)).
The diffuse satellite reflections obey reflection conditions resulting from internal translational symmetry, as well as from external translational symmetry.
The reflection conditions as listed in the \emph{International Tables for Crystallography C} \cite{bryan1993international} can be directly applied to the disordered superspace approach.

In Figure \ref{fig:DiffuseSatellite2d_2}(c) the simulated average structure is in the plane space group $pg$, with one atom at $(0.1,0)$ and the one atom on the symmetry related site at $(-0.1,0.5)$.
The modulation functions for the second site are transformed for the superspace group $pg1\left(\frac{1}{2}\beta\right)$ by the principles described in \cite{van2007incommensurate}.
The main Bragg reflections are extinct for $0k: \ k \ne 2n$.
In a modulated crystal structure, with an ordered superspace the satellite reflections that belong to the extinct main reflections are extinct as well.
The disordered superspace approach maintains the overall average symmetry and therefore the broad diffuse maxima obey the same reflection condition:
In Figure \ref{fig:DiffuseSatellite2d_2}(c) no diffuse maxima are observed at $01\pm \bm{q}$ as the parent Bragg reflection is extinct. At  $02\pm \bm{q}$ the diffuse maxima are observed as the parent Bragg reflection $02$ is present.

In Figure \ref{fig:DiffuseSatellite2d_2}(d) the simulated average structure is in the plane space group $pm$, with one atom at $(0.1,0)$ and the one atom on the symmetry related site at $(-0.1,0.0)$.
The main Bragg reflections show no extinction condition, as no symmetry elements containing translational elements are included in the plane space group descriptor.
An additional translational element along the internal superspace direction is introduced in the superspace group $pm1(0\beta)s0$.
In an ordered superspace all satellite reflections $0km: \ m\ne2n$ are extinct due to an intrinsic translation in superspace.
For the simulated case of occupational disorder only first order satellite reflections are observed, hence no satellite reflections are observed for parent Bragg reflections $0k$.
This extinction condition can be directly transferred to the diffuse maxima in the disordered superspace approach:
In Figure \ref{fig:DiffuseSatellite2d_2}(d) no diffuse maxima are observed at $0k\pm \bm{q}$. For all other parent Bragg reflections with $h\ne 0$ the broad diffuse satellites at $hk\pm \bm{q}$ are observed.

It is worth mentioning that the diffraction pattern shown in Figure \ref{fig:DiffuseSatellite2d_2}(d) is generated by a two dimensional crystal structure showing pure occupational disorder. 
The extinction of satellite reflections at $h=0$ would suggest a translational disorder with a displacement direction along $\bm{a}_{1}$.
The reflection conditions as described by superspace crystallography, call for caution in the direct interpretation of such diffuse scattering patterns and need a careful analysis of both, diffuse and Bragg scattering data to determine the underlying disorder.

By adding integer numbers to any of the components of the modulation wave vector the position of the diffuse maxima can be tuned in reciprocal space.
The disordered superspace approach allows a direct access to extinction conditions observed in single crystal diffuse scattering.

\section{Analytical expression for the disordered superspace approach}
This section will outline an expression for the diffuse scattering intensity from a primitive disordered superspace model, with one occupied site per unit-cell.
The analytical expression here is constrained to occupational modulations of the type:
\begin{equation}
p_{A}(\bm{x}) = m_{A} + A\cos(2\pi\bm{q}\bm{x})
\end{equation}
Here $m_{A}$ is the average probability to find an atom of type $A$.
The disordered superspace is composed of two types of modulation functions $P$ and $M$. 
The superspace composition is given as $m_{P} = 1-m_{M}$.

The probability to find an atom of type A on a site $\bm{x}$ depends on the type of modulation function at $(\bm{x},0)$ in the disordered superspace model.
The probability to find an atom of type A on a site $\bm{x}$ for the modulation function type $P$ is denoted $p_{A}^{M}(\bm{x})$. 
The probabilities $p_{A}^{P}(\bm{x})$, $p_{B}^{M}(\bm{x})$ and $p_{B}^{P}(\bm{x})$ are defined accordingly:
\begin{eqnarray}
p_{A}^{M}(\bm{x}) &= 	m_{A} - A\cos(2\pi\bm{q}\bm{x}) \\
p_{A}^{P}(\bm{x}) &= 	m_{A} + A\cos(2\pi\bm{q}\bm{x}) \\
p_{B}^{M}(\bm{x}) &= 	m_{B} + A\cos(2\pi\bm{q}\bm{x}) \\
p_{B}^{P}(\bm{x}) &= 	m_{B} - A\cos(2\pi\bm{q}\bm{x}) 
\end{eqnarray}
where the probabilities fulfil $p_{B}(\bm {x}) = 1 - p_{A}(\bm {x})$.

An analytical expression for the diffuse scattering intensity requires the definition of pair-probabilities.
Depending on the modulation functions on the site at $(\bm{x}_{i},0)$ and on the site at $(\bm{x}_{j},0)$, the probability to find a pair $AA$ varies. 
Here, the 16~different pair-probabilities of type $p^{PP}_{AA}(\bm{x}_{i},\bm{x}_{j})$ need to be calculated.
The expressions for $p^{PP}_{AA}(\bm{x}_{i},\bm{x}_{j})$, $p^{PM}_{AA}(\bm{x}_{i},\bm{x}_{j})$, $p^{MP}_{AA}(\bm{x}_{i},\bm{x}_{j})$ and $p^{MM}_{AA}(\bm{x}_{i},\bm{x}_{j})$ are stated here, the others can be calculated accordingly.
\begin{eqnarray}
p^{PP}_{AA}(\bm{x}_{i},\bm{x}_{j}) =& p_{A}^{P}(\bm{x}_{i}) \cdot p_{A}^{P}(\bm{x}_{j}) \nonumber \\
 =& m_{A}^{2} + m_{A}A\left[\cos(2\pi\bm{q}\bm{x}_{i}) + \cos(2\pi\bm{q}\bm{x}_{j})\right] \nonumber \\
& + A^{2}\cos(2\pi\bm{q}\bm{x}_{i})\cos(2\pi\bm{q}\bm{x}_{j})  \\
p^{PM}_{AA}(\bm{x}_{i},\bm{x}_{j})  =& m_{A}^{2} + m_{A}A\left[\cos(2\pi\bm{q}\bm{x}_{i}) - \cos(2\pi\bm{q}\bm{x}_{j})\right] \nonumber\\
 & - A^{2}\cos(2\pi\bm{q}\bm{x}_{i})\cos(2\pi\bm{q}\bm{x}_{j}) \\
p^{MP}_{AA}(\bm{x}_{i},\bm{x}_{j})  =& m_{A}^{2} - m_{A}A\left[\cos(2\pi\bm{q}\bm{x}_{i}) - \cos(2\pi\bm{q}\bm{x}_{j})\right] \nonumber\\
&- A^{2}\cos(2\pi\bm{q}\bm{x}_{i})\cos(2\pi\bm{q}\bm{x}_{j}) \\
p^{PP}_{AA}(\bm{x}_{i},\bm{x}_{j})  =& m_{A}^{2} - m_{A}A\left[\cos(2\pi\bm{q}\bm{x}_{i}) + \cos(2\pi\bm{q}\bm{x}_{j})\right] \nonumber\\
& + A^{2}\cos(2\pi\bm{q}\bm{x}_{i})\cos(2\pi\bm{q}\bm{x}_{j}) 
\end{eqnarray}

For the calculation of the diffuse scattering intensity, it is necessary to calculate the expectation value of the probability to find a pair of type $AA$ separated by a vector $\bm{v}$.
Therefore, $\langle p^{PP}_{AA}(\bm{v})\rangle$ needs to be calculated:
\begin{equation}
\begin{split}
\langle p^{PP}_{AA}(\bm{v})\rangle & = \frac{1}{N} \sum_{i=1}^{N} p^{PP}_{AA}(\bm{x}_{i},\bm{x}_{i} +\bm{v}) \\
& = m_{A}^{2} + \frac{A^{2}}{2}\cos(2\pi\bm{q}\bm{v}) \\
&+ \frac{1}{N} \sum_{i=1}^{N} \Bigl\{ m_{A}A\left[\cos(2\pi\bm{q}\bm{x}_{i}) + \cos(2\pi\bm{q}(\bm{x}_{i}+\bm{v}))\right] \Biggr.\\
& \Biggl. \qquad \qquad + \frac{A^{2}}{2} \cos(2\pi\bm{q}(2\bm{x}_{i}+
\bm{v})) \Bigr\} \\
&= m_{A}^{2} + \frac{A^{2}}{2}\cos(2\pi\bm{q}\bm{v}) 
\end{split}
\label{eq:p_pp_AA1}
\end{equation}
where $N$ is the number of atoms in the crystal. 
The remaining 15 pair probabilities can be calculated accordingly.

The probability to find a pair of modulation functions in superspace $PM$ separated by a vector $(\bm{v},0)$ is defined by the superspace Warren-Cowley short range order parameter:
\begin{equation}
\alpha^{s}_{\bm{v}} = 1- \frac{p^{PM}_{(\bm{v},0)}}{m_{P}m_{M}}
\label{eq:alpha_s}
\end{equation}
The diffuse scattering intensity can then be calculated, analogous to the formulas presented in \cite{schmidt2017diffuse}:
\begin{equation}
\begin{split}
I_{D}(\bm{h}) = N &\left| f_{A}(\bm{h}) - f_{B}(\bm{h})\right|^{2} \Biggl\{ 1+ \Biggr. \\
\Biggl.&\sum_{\bm{v} \in V_{p}} \left[(m_{P}-m_{M})^{2} + 4m_{P}m_{M}\alpha^{s}_{\bm{v}} \right]  A^{2} \Biggr. \\
& \qquad \quad \Biggl. \cos(2\pi\bm{q}\bm{v}) \cos(2\pi\bm{h}\bm{v}) \Biggr\}
\end{split}
\label{eq:ID_super}
\end{equation}

For a positive correlation ($\alpha^{s}_{\bm{v}}>0$), the diffuse scattering intensity shows diffuse maxima at $\bm{H}\pm\bm{q}$. 
The shape of the diffuse maximum is governed by the superspace correlations $\alpha^{s}_{\bm{v}}>0$.
For large inter-unit-cell superspace vectors $\bm{v}$ the superspace correlations $\alpha^{s}_{\bm{v}}\to0$.

For an unequal superspace composition $m_{P} \ne m_{M}$ even at large inter-unit-cell superspace vectors $\bm{v}$ the terms with $(m_{P}-m_{M})^{2} $ contribute to the diffuse scattering.
This gives rise to the sharp satellite reflections at $\bm{H}\pm\bm{q}$ on top of the diffuse scattering.

This analytical approach shows that the shape of the diffuse maxima in reciprocal space can be fully described by the Warren-Cowley short range order parameters $\alpha^{s}_{\bm{v}}$ in superspace, while the position is soley defined by the $\bm{q}$-vector.

\section{Possible applications}
Incommensurately modulated crystals are often encountered at phase transitions \cite{van2007incommensurate}.
Such a phase transition is an ordering process, in general from a disordered towards an ordered crystal structure upon lowering of the temperature.
In some cases a phase transition was observed, where diffuse scattering was observed above the phase transition and below a modulated crystal structure was observed (see e.g. \cite{folkers2018mystery}).
We suggest a disordered superspace model to describe such phase transitions.

Diffuse rods along high symmetry directions have been frequently observed in single crystal diffuse scattering (see \cite{withers2015modulation} and references therein).
The disordered superspace approach as demonstrated here allows the straight forward generation of such diffuse rods and also provides direct access to possible observed extinction conditions.

A superspace composition, which differs from $m_{P} = m_{M} = 0.5$ allows to generate sharp satellite reflections on top of broad diffuse maxima.
The mullite system as presented in \cite{klar2018exploiting} is only one example of a system showing sharp satellite reflections on top of diffuse scattering.

Another application of the proposed disordered superspace model is in the field of diffuse scattering where reverse Monte-Carlo modelling is used:
In reverse Monte-Carlo modelling a crystal structure is altered at random until the observed and calculated diffuse scattering overlap.
The diffuse scattering consists of broad maxima at fixed positions in reciprocal space. 
Using the proposed disordered superspace model a building principle can be directly derived from the position and width of the diffuse maximum.
The resulting structure has the same or even more physical meaning as the result of a reverse Monte-Carlo simulation but the computational cost is drastically reduced.
A direct interpretation of the diffuse maxima in terms of correlation parameters and modulation functions is possible.

In general the described disordered superspace model allows the generation of a crystal structure with a diffuse maximum at an arbitrary position in reciprocal space. 
The structures generated here show diffuse maxima at $\pm\bm{q}$ from each Bragg reflection. 
This may seem a restriction on the diffuse scattering, but considering Equation \ref{eq:SRO}, purely occupational disorder is also symmetric around Bragg reflections.
Hence for purely occupational disordered systems the disordered superspace model does not restrict possible configurations in reciprocal space.

\section{Conclusion}
A formalism is demonstrated that allows the interpretation of diffuse maxima as the result of a disordered superspace. 
The computational procedure demonstrated here allows the fast realization of structural models from reciprocal space analysis: 
The position and width of a diffuse maximum can be directly transferred into a disordered superspace model.
The superspace model only needs few parameters, namely the modulation wave vector and the Warren-Cowley short range order parameters for the ordering of the modulation functions in superspace, to sufficiently describe a disorder model.
The comparison to disorder description with correlation coefficients can be easily achieved, as the superspace approach delivers a recipe for structure building.
The disordered superspace approach combines the tools supplied by superspace crystallography and the description of pair correlations from conventional diffuse scattering analysis, to provide a more general and flexible description of diffuse maxima at arbitrary positions in reciprocal space.

The reasons listed above are why we suggest, that the disordered superspace model building, is superior to reverse Monte-Carlo modelling methods in diffuse scattering analysis. 
The reduction of disorder parameters as compared to correlation parameter driven description is immense and thus simplifies the interpretation of diffuse scattering.

\begin{acknowledgments}
E.M.S. thanks Bavarian Equal Opportunities Sponsorship -- Realisierung von Frauen in Forschung und Lehre (FFL) -- Realization Equal Opportunities for Women in Research and Teaching for funding.
P.B. Klar is acknowledged for the introduction to superspace crystallography and fruitful discussion.
\end{acknowledgments}

\appendix*
\section{Computational realization of a disordered superspace model}
The disordered superspace structures are simulated using the DISCUS program \cite{neder2008diffuse}.
A one-dimensional crystal of 20~000 unit-cells is generated.
Half of the atoms are of type $P$ (modulation function $p_{+}$) and half of the atom are of type $M$ (modulation function $p_{-}$).
Using a Monte Carlo algorithm the atoms are sorted with a positive correlation, yielding e.g. a desired positive correlation given by $\alpha^{s,1}_{1} = 0.85$,
This generates phase domains in the direction of $\bm{a}_{s,1}$.

The disordered modulated crystal structure is an AuAg 1:1 crystal, with lattice constant 5~\AA.
The atoms are placed at the origin of the unit-cell.
A loop over all atoms in the disordered superspace structure is taken.
If the modulation function is of type $p_{+}$, an Au atom is inserted in physical space at $x$ with probability:
\begin{equation}
p_{+}^{Au}(x) = 0.5 + 0.5\cos(2\pi qx)
\end{equation}
A evenly distributed random number in the interval $[0,1]$ is generated.
If this random number is larger than $p_{Au}(x)$ an Au atom is inserted, else an Ag atom is inserted.
If the modulation function is of type $p_{-}$, an Au atom is inserted in physical space with probability:
\begin{equation}
p_{-}^{Au}(x) = 0.5 - 0.5\cos(2\pi qx)
\end{equation}

This procedure generates a physical structure with equal probabilities for Au and Ag atoms.
The main Bragg reflections were omitted in the calculation of the diffraction pattern.
The approach was extended for the $(2+1)$-dimensional diffraction patterns shown in Figure \ref{fig:DiffuseSatellite2d_2}.

%

\bibliography{SuperSpaceDisorder_arxiv}

\end{document}